\documentclass[lettetsize,journal]{IEEEtran}
\usepackage{amsmath,amsfonts}
\usepackage{algorithmic}
\usepackage{algorithm}
\usepackage{array}
\usepackage{textcomp}
\usepackage{stfloats}
\usepackage{url}
\usepackage{verbatim}
\usepackage{graphicx}
\usepackage[caption=false,font=footnotesize]{subfig}
\usepackage{booktabs}
\usepackage{bm}
\usepackage{diagbox}
\usepackage{xcolor}
\usepackage{colortbl}
\usepackage{threeparttable}
\usepackage{cite}
\usepackage[hidelinks]{hyperref}
\begin{document}

\title{Channel Estimation for RIS-Assisted mmWave Systems \\ via Diffusion Models}

\author{Yang~Wang,~\IEEEmembership{}
        Yin~Xu,~\IEEEmembership{}
        Cixiao~Zhang,~\IEEEmembership{}
        Zhiyong~Chen,~\IEEEmembership{}
        Mingzeng~Dai,~\IEEEmembership{}
        Haiming~Wang,~\IEEEmembership{}
        Bingchao~Liu,~\IEEEmembership{}
        
        Dazhi~He,~\IEEEmembership{}
        and~Meixia~Tao,~\IEEEmembership{Fellow,~IEEE,}
        
\thanks{This work was supported in part by the Shanghai Municipal Education Commission under Grant 2024AIYB002, and in part by the National Natural Science Foundation of China Program under Grant 62422111, Grant 62271316, Grant 62371291, Grant 62431015, and Grant 62222111.}
\thanks{Yang Wang, Yin Xu (\textit{Senior Member, IEEE}), Cixiao Zhang, Zhiyong Chen (\textit{Senior Member, IEEE}), Dazhi He (\textit{Senior Member, IEEE}), and Meixia Tao (\textit{Fellow, IEEE}) are with the Cooperative Medianet
Innovation Center, Shanghai Jiao Tong University, Shanghai 200240, China
(e-mail: \{vy\_eddie, xuyin, cixiaozhang, zhiyongchen, hedazhi, mxtao\}@sjtu.edu.cn).}
\thanks{Mingzeng Dai, is with the Lenovo Group, Lenovo Research, Shanghai 201203, China (e-mail: daimz4@lenovo.com). Haiming Wang and Bingchao Liu are with the Lenovo Group, Lenovo Research, Beijing 100094, China (e-mail: \{wanghm14, liubc2\}@lenovo.com).}
\thanks{The corresponding author is Yin Xu.}
}



\maketitle

\begin{abstract}
Reconfigurable intelligent surface (RIS) has been recognized as a promising technology for next-generation wireless communications. However, the performance of RIS-assisted systems critically depends on accurate channel state information (CSI). To address this challenge, this letter proposes a novel channel estimation method for RIS-aided millimeter-wave (mmWave) systems based on diffusion models (DMs). Specifically, the forward diffusion process of the original signal is formulated to model the received signal as a noisy observation within the framework of DMs. Subsequently, the channel estimation task is formulated as the reverse diffusion process, and a sampling algorithm based on denoising diffusion implicit models (DDIMs) is developed to enable effective inference. Furthermore, a lightweight neural network, termed BRCNet, is introduced to replace the conventional U-Net, significantly reducing the number of parameters and computational complexity. Extensive experiments conducted under various scenarios demonstrate that the proposed method consistently outperforms existing baselines.
\end{abstract}


\section{Introduction}
\IEEEPARstart{M}{illimeter-wave} (mmWave) communication is a key technology for the sixth generation (6G) communication systems to fully exploit the available spectrum \cite{Wang2023}. Nevertheless, the coverage range of mmWave signals is limited due to their rapid attenuation, which makes them susceptible to blockage. To overcome this limitation, reconfigurable intelligent surfaces (RISs) \cite{Wu2019} have been introduced to extend communication coverage and improve transmission rates. However, the effectiveness of the RIS depends on the acquisition of accurate channel state information (CSI). In general, a RIS consists of passive elements without signal processing capabilities, which poses a significant challenge for channel estimation in RIS-assisted systems.

Conventional methods (e.g., linear minimum mean square error, LMMSE) and compressed sensing (CS)-based methods (e.g., the double-structured orthogonal matching pursuit, DS-OMP~\cite{Wei2021}) struggle to achieve accurate channel estimation under low signal-to-noise ratio (SNR) conditions. With the rapid development of deep learning (DL), DL-based methods have been increasingly applied in RIS-assisted systems. For instance, a deep residual network (DRN) is used to denoise the results obtained from the least squares (LS) estimator in~\cite{Liu2021}. Furthermore, a global attention residual network (GARN) is introduced to fuse multi-channel information and improve the accuracy of channel estimation~\cite{Feng2023}. However, these direct-mapping DL methods require network retraining for different SNRs, limiting their generalizability and practicality.

Recently, diffusion models (DMs)~\cite{Ho2020} have garnered significant attention in wireless communications due to their strong capability to model complex data distributions~\cite{Wu2024, Guan2024, Zhou2025, Fesl2024}. Among their various applications, channel estimation has received particular attention due to its critical role, especially in RIS-assisted scenarios where cascaded and direct links lead to more complex distributions and greater challenges. In~\cite{Tong2024}, the authors employ received signals to guide the sampling of DMs for channel estimation. Another approach has been proposed in~\cite{Zhang2025}, which infers the full channel from the partial channel based on conditional DMs. Moreover, in~\cite{Liu2022}, the diffusion process is conducted in the spectral domain to improve stability. However, in these works, the sampling process typically starts from the maximum step $T$ without adapting to different SNR conditions, which limits their robustness under varying channel environments. Furthermore, the network architectures adopted in previous works fail to strike a balance among performance, parameter size, and computational complexity. Therefore, achieving accurate and efficient channel estimation in RIS-assisted systems remains a challenging problem that requires further exploration.

In this letter, we propose a novel and practical channel estimation method for RIS-assisted mmWave systems based on DMs. The key contributions can be summarized as follows.

\begin{itemize}
\item We formulate both direct and cascaded channel estimation in RIS-assisted scenarios within the framework of denoising diffusion implicit models (DDIMs), and introduce a step-matching mechanism that allows the estimation process to adapt to different SNR conditions.

\item We initially adopt a U-Net-based channel estimation scheme as a baseline, and subsequently propose a parameter-efficient architecture, termed BRCNet, achieving an improved trade-off among estimation performance, model complexity, and computational overhead.

\item Extensive simulations demonstrate that the proposed method outperforms the baselines under various scenarios, while the proposed BRCNet achieves performance comparable to U-Net using only 9.34\% of its parameters.
\end{itemize} 

\IEEEpubidadjcol

\section{System Model}
\label{sec:System Model}
\subsection{Channel Model}
\label{sec:Channel Model}

\begin{figure}[!t]
\centering
\includegraphics[width=3.4in]{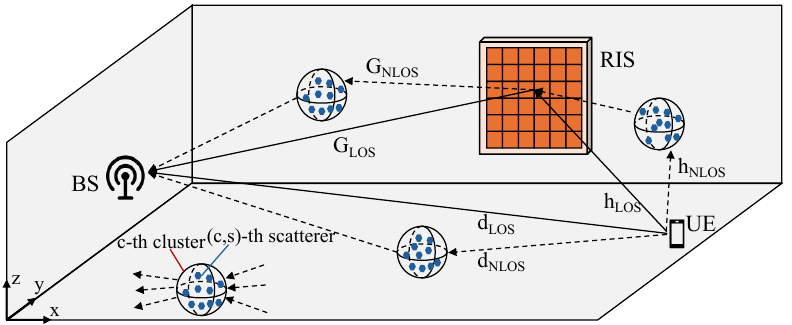}
\caption{RIS-assisted mmWave communication systems.}
\label{Fig_1}
\end{figure}

As illustrated in Fig. \ref{Fig_1}, we consider the uplink channel estimation for RIS-assisted mmWave systems. Both the base station (BS) and RIS are equipped with uniform planar arrays (UPAs) with half-wavelength element spacing, while the user equipment (UE) has a single antenna. The BS has $M=M_h \times M_v$ antennas, and the RIS consists of $N=N_h \times N_v$ passive elements. The UE-BS, UE-RIS, and RIS-BS channels are denoted as $\mathbf{d} \in \mathbb{C}^{M \times 1}$, $\mathbf{h} \in \mathbb{C}^{N \times 1}$, and $\mathbf{G} \in \mathbb{C}^{M \times N}$, respectively. Following \cite{3GPP2018} and \cite{Basar2021}, a clustered statistical channel model is adopted. The RIS-BS channel consists of line-of-sight (LOS) and non-line-of-sight (NLOS) components, i.e., $\mathbf{G} = \mathbf{G}_{\text{LOS}} + \mathbf{G}_{\text{NLOS}}$. These two components are represented as

\begin{equation}
    \label{eq_1}
    \begin{aligned}
        \mathbf{G}_\text{LOS} =& 
        \sqrt{G_e\left( \varphi^{G_t}_{\text{LOS}} \right)  L_{\text{RIS-BS}}}
        \mathbf{a}_\text{BS}\left( \phi^{G_r}_{\text{LOS}}, \varphi^{G_r}_{\text{LOS}} \right) \\
        &\times
        \mathbf{a}^T_\text{RIS}\left( \phi^{G_t}_{\text{LOS}}, \varphi^{G_t}_{\text{LOS}} \right),
    \end{aligned}
\end{equation}

\begin{equation}
    \label{eq_2}
    \begin{aligned}
        \mathbf{G}_\text{NLOS} =& 
        \gamma^G \sum_{c=1}^{C^G} \sum_{s=1}^{S^G_c} 
        \beta^G_{c,s} 
        \sqrt{G_e\left( \varphi^{G_t}_{c,s} \right) L_{c,s}^{G_t}}
        \mathbf{a}_\text{BS}\left( \phi^{G_r}_{c,s}, \varphi^{G_r}_{c,s} \right) \\
        &\times \mathbf{a}_\text{RIS}^T\left( \phi^{G_t}_{c,s}, \varphi^{G_t}_{c,s} \right),
    \end{aligned}
\end{equation}

\noindent where $C^{G}$ and $S^G_c$ denote the number of clusters and the number of scatterers in the $c$-th cluster, respectively. The normalization factor is defined as $\gamma^G=\sqrt{ \frac{1}{\sum_{c=1}^{C^G} S_c^G} }$. $\beta^G_{c,s} \sim \mathcal{CN}(0,1)$ denotes the path gain of the $(c,s)$-th scatterer, which follows a circularly symmetric complex Gaussian (CSCG) distribution. The element radiation pattern of the RIS corresponding to the $(c,s)$-th scatterer is defined as $G_e\left( \varphi^{G_t}_{c,s} \right) = 2(2\zeta+1)\cos^{2\zeta}(\varphi^{G_t}_{c,s})$. The path loss $L_{c,s}^{G_t}$ is defined as

\begin{equation}
\label{eq_3}
    \begin{aligned}
        L_{c,s}^{G_t} =& 
        -20 \log_{10} \left( \frac{4\pi}{\lambda} \right) \\
        &-10n \left( 1 + b \left( \frac{f_c - f_0}{f_0} \right) \right) 
        \log_{10}(d_{c,s}) - X_{\sigma_x},
    \end{aligned}
\end{equation}

\noindent where $\lambda$ is the carrier wavelength, $n$ is the path loss exponent, $b$ is a system parameter,  $f_0$ is the reference frequency, $f_c$ is the carrier frequency, $d_{c,s}$ is the propagation distance of the $(c,s)$-th scatterer, and $X_{\sigma_x} \sim \mathcal{N}(0,\sigma_x^2)$ is the shadow factor.

$\mathbf{a}_\text{BS}\left( \phi^{G_r}_{c,s}, \varphi^{G_r}_{c,s} \right)$ and $\mathbf{a}_\text{RIS}\left( \phi^{G_t}_{c,s}, \varphi^{G_t}_{c,s} \right)$ denote the steering vectors of the BS and the RIS, respectively. Here, $\phi^{G_r}_{c,s}$ and $\varphi^{G_r}_{c,s}$ represent the azimuth and elevation angles of arrival (AOA), respectively, while $\phi^{G_t}_{c,s}$ and $\varphi^{G_t}_{c,s}$ denote the corresponding angles of departure (AOD). $\phi^{G_t}_{c,s}$ follows a Laplacian distribution $\mathcal{L}\left( \phi^{G_t}_c, \sigma_{\phi} \right)$, where $\phi^{G_t}_c$ follows a uniform distribution $\mathcal{U}\left[-\pi/2,\pi/2\right]$ and $\sigma_\phi$ represents the angular spread. Similarly, $\varphi^{G_t}_{c,s}$ follows $ \mathcal{L}\left( \varphi^{G_t}_c, \sigma_{\varphi} \right)$ with $\varphi^{G_t}_c \sim \mathcal{U}\left[-\pi/4,\pi/4\right]$. The UPA steering vectors can be written as

\begin{equation}
\label{eq_4}
    \begin{aligned}
        \mathbf{a}(\phi, \varphi) = 
        [
        &1, \ldots, 
        e^{j \frac{2\pi d}{\lambda} (x \sin\varphi + y \sin\phi \cos\varphi)}, \\ &\ldots, 
        e^{j \frac{2\pi d}{\lambda} \left( (N_h - 1) \sin\varphi + (N_v - 1) \sin\phi \cos\varphi \right)}
        ]^T,
    \end{aligned}
\end{equation}

\noindent where $0 \leq x \leq N_h - 1$ and $0 \leq y \leq N_v - 1$, and $d$ denotes the element spacing. Similarly, the UE-RIS channel $\mathbf{h}$ and the UE-BS channel $\mathbf{d}$ share the same structural form as $\mathbf{G}$.

\subsection{Signal Model}
The received signal $\mathbf{y}_{k} \in \mathbb{C}^{M \times 1}$ in the $k$-th time slot can be expressed as

\begin{equation}
    \label{eq_5}
    \mathbf{y}_k = \left( \mathbf{d} + \mathbf{G} \operatorname{diag}(\bm{\omega}_{k}) \mathbf{h}\right) x_k + \mathbf{n},
\end{equation}

\noindent where $ \mathbf{n} \sim \mathcal{CN}(0,2\sigma^{2}\mathbf{I})$ denotes the additive white Gaussian noise (AWGN), $x_k$ is the pilot symbol, and $\operatorname{diag}(\cdot)$ denotes the  diagonal operator. The phase shift vector is given by $ \bm{\omega}_{k} = [\beta_1 e^{j \theta_{1}}, \beta_2 e^{j \theta_{2}},\dots,\beta_N e^{j \theta_{N}}]^T$, where $\beta_n \in [0, 1]$ and $\theta_n \in [-\pi,\pi)$ represent the amplitude and the phase shift of the $n$-th element, respectively. The channel of the reflecting link UE-RIS-BS can be rewritten as $\mathbf{G} \operatorname{diag}(\bm{\omega}_{k}) \mathbf{h} = \mathbf{G} \operatorname{diag}(\mathbf{h})\bm{\omega}_{k}$. We define the cascaded channel as $\mathbf{H}_c=\mathbf{G} \operatorname{diag}(\mathbf{h})$ and the augmented phase shift vector $\mathbf{p}_k=[1, \bm{\omega}_k^T]^T \in \mathbb{C}^{(N+1) \times 1}$. Based on (\ref{eq_5}), the received signal can be expressed as

\begin{equation}
    \label{eq_6}
    \mathbf{y}_k = \mathbf{H} \mathbf{p}_k x_k + \mathbf{n},
\end{equation}

\noindent where $\mathbf{H} = \left[ \mathbf{d},\ \mathbf{H}_c \right] \in \mathbb{C}^{M \times (N+1)}
$. Without loss of generality, the pilot symbol is set to $x_k=1$ \cite{Feng2023, Tong2024}. After $K$ time slots, the received signal $\mathbf{Y} \in \mathbb{C}^{M \times K}$ is given by

\begin{equation}
    \label{eq_7}
    \mathbf{Y}=\mathbf{H} \mathbf{P} + \mathbf{N}.
\end{equation}

The RIS reflecting matrix is set as the columns of the discrete Fourier transform (DFT) matrix \cite{Zheng2020}. The objective of channel estimation is to recover the unknown channel $\mathbf{H}$ from the received signal $\mathbf{Y}$. By vectorizing \eqref{eq_7}, we obtain

\begin{equation}
\label{eq_8}
    \text{vec}(\mathbf{Y})=(\mathbf{I}_K \otimes \mathbf{H})\cdot \text{vec}(\mathbf{P}) + \text{vec}(\mathbf{N}),
\end{equation}

\noindent where $\operatorname{vec}(\cdot)$ denotes vectorization and $\otimes$ denotes the Kronecker product. Given the noise power $\sigma^2$, \eqref{eq_8} is normalized and converted into an equivalent real-valued form as

\begin{equation}
    \label{eq_9}
   \mathbf{y} = \frac{1}{\sqrt{1+\sigma^{2}}} \mathbf{A} \mathbf{p} + \frac{\sigma}{\sqrt{1+\sigma^{2}}}\mathbf{n},
\end{equation}

\noindent where $\mathbf{y} \in \mathbb{R}^{2MK \times 1}$, $\mathbf{A} \in \mathbb{R}^{2MK \times 2(N+1)K}$, $\mathbf{p} \in \mathbb{R}^{2(N+1)K \times 1}$, and $\mathbf{n} \in \mathbb{R}^{2MK \times 1}$ with entries distributed as $n_{i} \sim \mathcal{N}(0,1)$.

\section{Proposed Method}

This section details the proposed method. First, a forward diffusion process for the received signal is formulated to define the training procedure. Then, a sampling algorithm based on DDIMs is developed. Furthermore, a lightweight neural network architecture, termed BRCNet, is designed to reduce model complexity and computational overhead.

\subsection{Training Algorithm of the Proposed Method}

For clarity, the original signal in the diffusion process is defined as $\mathbf{x}_0 = \mathbf{A} \mathbf{p}$. By progressively adding noise to $\mathbf{x}_0$, the forward diffusion process simulates the effect of varying noise levels on the pilot signal after RIS-assisted transmission. Consequently, the signal at step $t \in \{1,2,...,T\}$ is given by

\begin{equation}
    \label{eq_10}
    \mathbf{x}_t = \sqrt{1 - \beta_t} \mathbf{x}_{t-1} + \sqrt{\beta_{t}} \boldsymbol{\epsilon}=\sqrt{\bar{\alpha}_t} \mathbf{x}_0 + \sqrt{1 - \bar{\alpha}_t} \boldsymbol{\epsilon},
\end{equation}

\noindent where $\alpha_{t}=1-\beta_{t}$, $\bar{\alpha}_t = \prod_{s=1}^{t} \alpha_s$, and $\bm{\epsilon} \sim \mathcal{N}(\mathbf{0},\mathbf{I})$ denotes Gaussian noise. The conditional distribution of $\mathbf{x}_t$ given $\mathbf{x}_0$ is

\begin{equation}
\label{eq_11}
    q(\mathbf{x}_t | \mathbf{x}_0) \sim \mathcal{N} \left( \mathbf{x}_t; \sqrt{\bar{\alpha}_t} \mathbf{x}_0,(1-\bar{\alpha}_t) \mathbf{I} \right).
\end{equation}

The objective of the diffusion model is to learn the distribution of $\mathbf{x}_0$, enabling accurate reconstruction of clean pilot responses from noisy observations. Following the DMs framework~\cite{Ho2020}, minimizing the variational bound leads to a noise-prediction objective, which is given by

\begin{equation}
\label{eq_12}
    \begin{aligned}
        &\mathbb{E} \left[ -\log p_{\theta} (\mathbf{x}_0) \right] 
        \leq \mathbb{E}_q \left[ -\log \frac{p_{\theta} (\mathbf{x}_{0:T})}{q(\mathbf{x}_{1:T} | \mathbf{x}_0)} \right] \\
        &= \mathbb{E}_q [ 
        \underbrace{D_{\text{KL}}( q(\mathbf{x}_T | \mathbf{x}_0) \parallel p(\mathbf{x}_T))}_{L_T} - \underbrace{\log p_{\theta} (\mathbf{x}_0 | \mathbf{x}_1)}_{L_0} \\
        &+ \sum_{t > 1} \underbrace{D_{\text{KL}}( q(\mathbf{x}_{t-1} | \mathbf{x}_t, \mathbf{x}_0) \parallel p_{\theta} (\mathbf{x}_{t-1} | \mathbf{x}_t))}_{L_{t-1}} ],
    \end{aligned}
\end{equation}

\noindent where $D_{\text{KL}}(\cdot)$ denotes the Kullback-Leibler (KL) divergence. The third term captures the gap between the true backward transition and its parameterized approximation. By aligning the parameterized transition with the true conditional distribution of the noisy signal, the model can achieve more accurate denoising and signal reconstruction in the RIS-assisted channel scenario. $q(\mathbf{x}_{t-1} | \mathbf{x}_t, \mathbf{x}_0)$ can be derived via Bayes' rule as in~\cite{Ho2020}. With a specific parameterization of $p_{\theta} (\mathbf{x}_{t-1} | \mathbf{x}_t)$, $L_{t-1}$ can be calculated in a Rao-Blackwellized fashion with closed-form expressions~\cite{Ho2020}. To summarize, the neural network's loss function can be formulated as

\begin{equation}
\label{eq_13}
L(\theta)=\mathbb{E}_{\mathbf{x}_0,\epsilon,t}\left[\| \boldsymbol{\epsilon} - \boldsymbol{\epsilon}_{\theta}(\mathbf{x}_t, t) \|^2 \right].
\end{equation}

By optimizing~\eqref{eq_13}, the network learns to estimate the noise component in $\mathbf{x}_{t}$, enabling progressive reconstruction of the clean signal during sampling. The inherent angular sparsity of RIS-assisted mmWave channels aligns well with the progressive denoising nature of diffusion models, facilitating more effective learning of the channel structure. Therefore, we adopt the angular-domain representation as in~\cite{Fesl2024}, and construct the training dataset as $\mathcal{X} = \left\{ \mathbf{X}_n = \operatorname{fft}(\mathbf{H}_n \mathbf{P}) \right\}_{n=1}^{N_{\mathrm{train}}}$, where $\operatorname{fft}(\cdot)$ denotes the two-dimensional fast Fourier transform (FFT). In summary, the training algorithm is outlined in Algorithm \ref{alg:alg1}\footnote{Consistent with \cite{Ho2020} and \cite{Fesl2024}, the vectorized form is employed for theoretical analysis, whereas the matrix form is used in practical implementations.}.

\begin{algorithm}[!t]
    \caption{Training Algorithm}
    \label{alg:alg1}
    \renewcommand{\algorithmicrequire}{\textbf{Input:}}
    \renewcommand{\algorithmicensure}{\textbf{Output:}}
    \begin{algorithmic}[1] 
        \REQUIRE Training set $ \mathcal{X} $, hyper-parameter $T$ and $\{\beta_t\}_{t=1}^{T}$
        \ENSURE Trained model $\boldsymbol{\epsilon}_{\theta}(\mathbf{x}_t, t)$
        \REPEAT
            \STATE Sample $\mathbf{X}$ from $\mathcal{X}$;
            \STATE Sample $t$ uniformly from $\{1,\dots, T\}$;
            \STATE Compute $\mathbf{X}_t$ based on \eqref{eq_10};
            \STATE Take gradient descent step according to \eqref{eq_13};
        \UNTIL{converged}
    \end{algorithmic}
\end{algorithm}

\begin{algorithm}[!t]
    \caption{Sampling Algorithm}
    \label{alg:alg2}
    \renewcommand{\algorithmicrequire}{\textbf{Input:}}
    \renewcommand{\algorithmicensure}{\textbf{Output:}}
    \begin{algorithmic}[1]
        \REQUIRE Received signal $\mathbf{Y}$, starting step index $t_s$
        \ENSURE $\hat{\mathbf{H}}$
        \STATE $\mathbf{X}_{t_s}=\operatorname{fft}(\frac{1}{\sqrt{1+\sigma^{2}}}\mathbf{Y})$;
        \FOR{$m=t_s,\dots, 2$}
            \STATE Compute $\mathbf{X}_{m-1}$ according to \eqref{eq_14};
        \ENDFOR
        \STATE $t=1$;
        \STATE $\mathbf{X}_{0} = \frac{1}{\sqrt{\alpha_1}} \left( \mathbf{X}_1 - \sqrt{1 - \bar{\alpha}_1} \boldsymbol{\epsilon}_{\theta}(\mathbf{X}_1, 1) \right)$;
        \STATE $\hat{\mathbf{H}} = \operatorname{ifft}(\mathbf{X}_0) \mathbf{P}^\dagger$;
    \end{algorithmic}
\end{algorithm}
    
\subsection{Sampling Algorithm of the Proposed Method}

During the sampling process, the proposed method progressively denoises the signal through the sequential transitions $\mathbf{x}_t \rightarrow \mathbf{x}_{t-1} \rightarrow \cdots \rightarrow \mathbf{x}_0$. This stepwise refinement restores the dominant angular components of the RIS-assisted channel from noisy observations, leading to more accurate channel estimation. According to \cite{Song2021}, $q(\mathbf{x}_{t-1} | \mathbf{x}_t, \mathbf{x}_0)$ is approximated by $p_{\theta}(\mathbf{x}_{t-1} | \mathbf{x}_t, \hat{\mathbf{x}}_0)$, yielding the following sampling process

\begin{equation}
\label{eq_14}
    \begin{aligned}
    \hat{\mathbf{x}}_{t-1} &= 
    \sqrt{ \bar{\alpha}_{t-1} } \underbrace{\left(
     \frac{ \hat{\mathbf{x}}_{t}
    - \sqrt{1 - \bar{\alpha}_{t}}\, \epsilon_\theta (\hat{\mathbf{x}}_{t}, t)  }
    { \sqrt{ \bar{\alpha}_{t} } } \right)
    }_{\text{estimate of } \mathbf{x}_0}\\
    &+ 
    \underbrace{
    \sqrt{1 - \bar{\alpha}_{t-1}}\, \epsilon_\theta (\hat{\mathbf{x}}_{t}, t)
    }_{\text{direction pointing to } \mathbf{x}_t}.
    \end{aligned}
\end{equation}

Given \eqref{eq_9} and \eqref{eq_10}, when $\sqrt{\bar{\alpha}_{t_s}} = \frac{1}{\sqrt{1+\sigma^{2}}}$, the KL divergence satisfies $D_{\mathrm{KL}}( q(\mathbf{x}_{t_s} |  \mathbf{x}_0) \parallel p (\mathbf{y} | \mathbf{x}_0)) = 0$. This indicates that the forward process yields signals that follow the same distribution as the received signals. Here, $\mathbf{x}_{t_s}$ denotes the signal at step $t_s$ in the forward process. Therefore, during the sampling process, we can start from an intermediate step $t_s$ rather than from step $T$ as done in \cite{Ho2020}, which significantly improves the efficiency. The index $t_s$ is computed as

\begin{equation}
\label{eq_15}
    t_s = \underset {t} {\arg \min} \left| 2\sigma^2 - \frac{1 - \bar{\alpha}_{t}}{\bar{\alpha}_{t}} \right|,
\end{equation}

\noindent which is referred to as the step-matching mechanism~\cite{Wu2024}. The starting step depends on the noise level, with larger indices indicating stronger noise contamination and thus requiring more denoising steps. By incorporating \eqref{eq_15}, the proposed approach can adapt to varying noise conditions without retraining.

In summary, the sampling process starts from the received signal, determines the starting step index according to \eqref{eq_15}, and repeatedly applies \eqref{eq_14} to obtain the denoised original signal $\mathbf{X}_0$. The channel estimation result $\hat{\mathbf{H}}$ is then obtained by applying the inverse fast Fourier transform, denoted as $\operatorname{ifft}(\cdot)$, to the denoised signal $\mathbf{X}_0$, and subsequently multiplying the result by the pseudo-inverse of $\mathbf{P}$, denoted as $\mathbf{P}^\dagger$. The proposed sampling algorithm is outlined in Algorithm \ref{alg:alg2}.

\subsection{Network Architecture Design}

\begin{figure}[!t]
\centering
\includegraphics[width=3.4in]{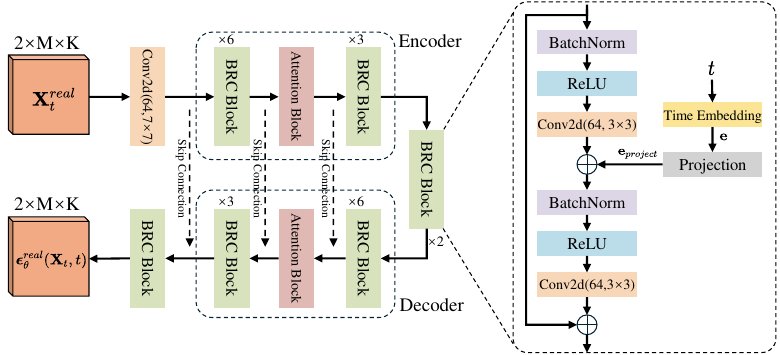}
\caption{The architecture of the proposed network.}
\label{Fig_2}
\end{figure}

In this subsection, we present a detailed description of the proposed BRCNet architecture, as illustrated in Fig. \ref{Fig_2}.

\textit{1) Preprocess:} To facilitate network processing, the complex-valued matrix $\mathbf{X}_t \in \mathbb{C}^{M \times K}$ is first decomposed into its real part $\Re(\mathbf{X}_t)$ and imaginary part $\Im(\mathbf{X}_t)$, and then reshaped into a three-dimensional tensor $\mathbf{X}^{\mathrm{real}}_t      \in \mathbb{R}^{2 \times M \times K}$.

\textit{2) $\text{Conv2d}(64, 7\times7)$:} Convolutional layers have been widely adopted in image processing tasks due to their ability to capture local correlations between pixels. Since both channel matrices and images can be represented in matrix form, we apply a convolutional layer with 64 output channels and a 7$\times$7 kernel, denoted as $\text{Conv2d}(64, 7\times7)$, to effectively extract correlations among elements within the input. 

\textit{3) Encoder:}
Motivated by the architectural efficiency and denoising effectiveness of DnCNN \cite{Zhang2017}, we adapt its design principles to channel estimation and introduce a novel BRC block. Specifically, this block consists of \textbf{B}atch normalization (BN) layers, \textbf{R}eLU activation functions, and \textbf{C}onvolutional layers to extract multi-level features efficiently. To further enhance the model’s capability, we integrate the self-attention mechanism. This tailored design not only improves denoising performance but also reduces model complexity, making it well-suited for practical wireless communication scenarios.
 
Temporal information $t$ is embedded into each BRC block via a sinusoidal positional embedding vector $\mathbf{e} \in \mathbb{R}^{N_{t_{embed}}}$, enabling the model to capture temporal dependencies. The embedding vector is then projected to $\mathbf{e}_{project} \in \mathbb{R}^{N_{project}}$ to facilitate subsequent computations by a projection module consisting of a Swish activation function and a linear layer.

\textit{4) Decoder:} The decoder adopts a structure similar to that of the encoder. To further extract hierarchical features, skip connections are employed to fuse multi-level information by concatenating feature maps from corresponding layers.

\textit{5) Postprocess:} The model outputs the estimated noise $\boldsymbol{\epsilon}^{real}_{\theta}(\mathbf{X}_t, t) \in \mathbb{R}^{2 \times M \times K}$, which is subsequently transformed into a complex-valued matrix $\boldsymbol{\epsilon}_{\theta}(\mathbf{X}_t, t) \in \mathbb{C}^{M \times K}$ by reversing the preprocessing steps applied to the model input.

\section{Numerical Results}
\subsection{Experiment Setup}

We evaluate the proposed method under four scenarios, combining two carrier frequencies (28~GHz and 73~GHz) with two propagation environments: Indoor Hotspot (InH) and Urban Microcell (UMi). The normalized mean square error (NMSE), defined as $    \mathrm{NMSE} = \mathbb{E} [  || \hat{\mathbf{H}} - \mathbf{H} ||_F^2 / || \mathbf{H} ||_F^2  ]$, is adopted as the evaluation metric. The coordinates $(x,y,z)$ of the BS and RIS (in meters) are set to $(0,25,2)$ and $(40,50,2)$ for InH, and $(0,25,20)$ and $(70,85,10)$ for UMi. Users are positioned at a fixed height of 1 m, with their horizontal locations randomly distributed within an 8 m-radius circle centered at the RIS. The channel parameters in Section~\ref{sec:Channel Model} follow the settings provided in~\cite{Basar2021}. Notably, in the InH scenario, the RIS is assumed to be close enough to the BS that the RIS–BS channel contains only the LOS component~\cite{Basar2021}, highlighting the distinct characteristics of different scenarios.

Following the configuration in \cite{Ho2020}, we set the total diffusion steps to $T=1000$. The variance schedule $\{\beta_{t}\}$ increases linearly from $\beta_{0}=1 \times 10^{-4}$ to $\beta_{T}=0.02$. Our dataset generation code is developed based on the open-source MATLAB simulator SimRIS~\cite{Basar2021}. For each scenario, the training set consists of 50,000 channel samples, while the validation and test sets each contain 10,000 channel samples. For training, the Adam optimizer is employed with a cosine warm-up learning rate schedule, where the initial learning rate is set to $1 \times 10^{-4}$. The model is trained for 800 epochs with a batch size of 256.

\subsection{Results and Analysis}

\begin{figure*}[!t]
    \centering
    \subfloat[UMi, $f_c=28$GHz]{%
        \includegraphics[width=0.24\textwidth]{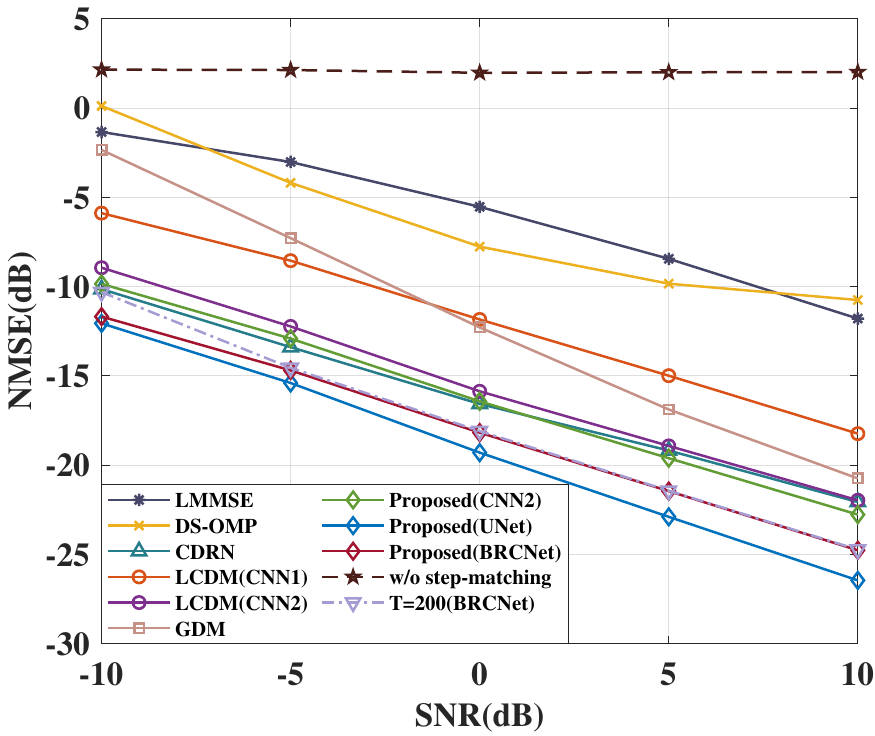}%
        \label{fig3:a}
    }
    \hfill
    \subfloat[UMi, $f_c=73$GHz]{%
        \includegraphics[width=0.24\textwidth]{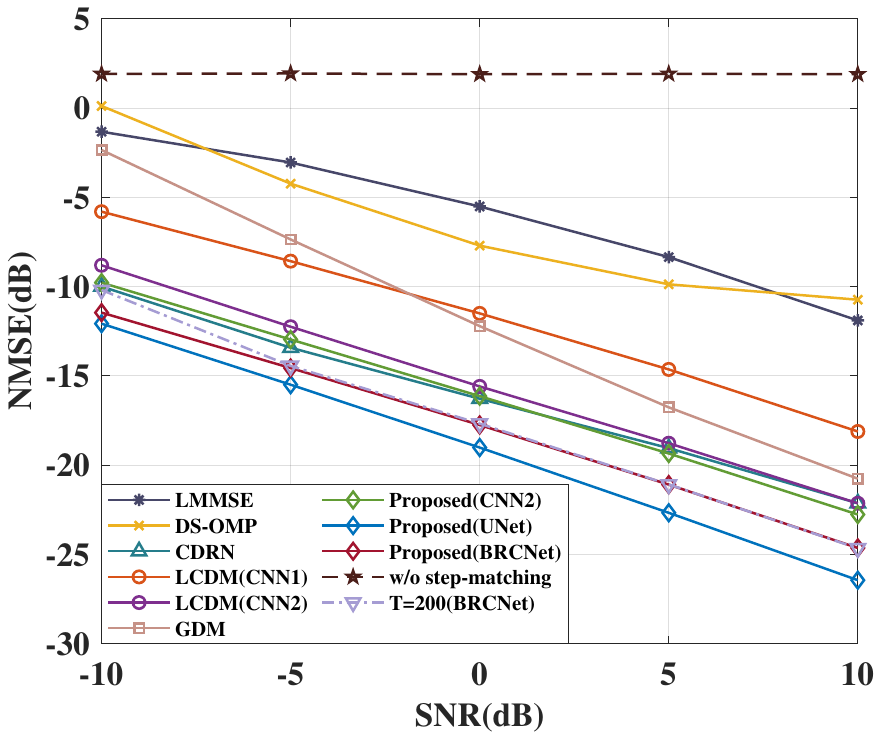}%
        \label{fig3:b}
    }
    \hfill
    \subfloat[InH, $f_c=28$GHz]{%
        \includegraphics[width=0.24\textwidth]{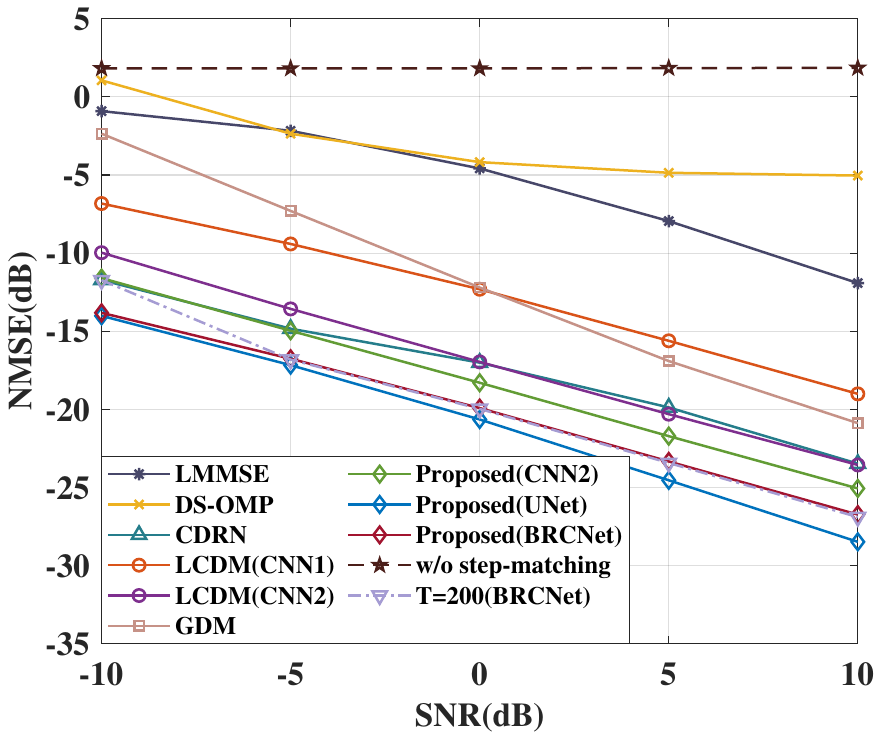}%
        \label{fig3:c}
    }
    \hfill
    \subfloat[InH, $f_c=73$GHz]{%
        \includegraphics[width=0.24\textwidth]{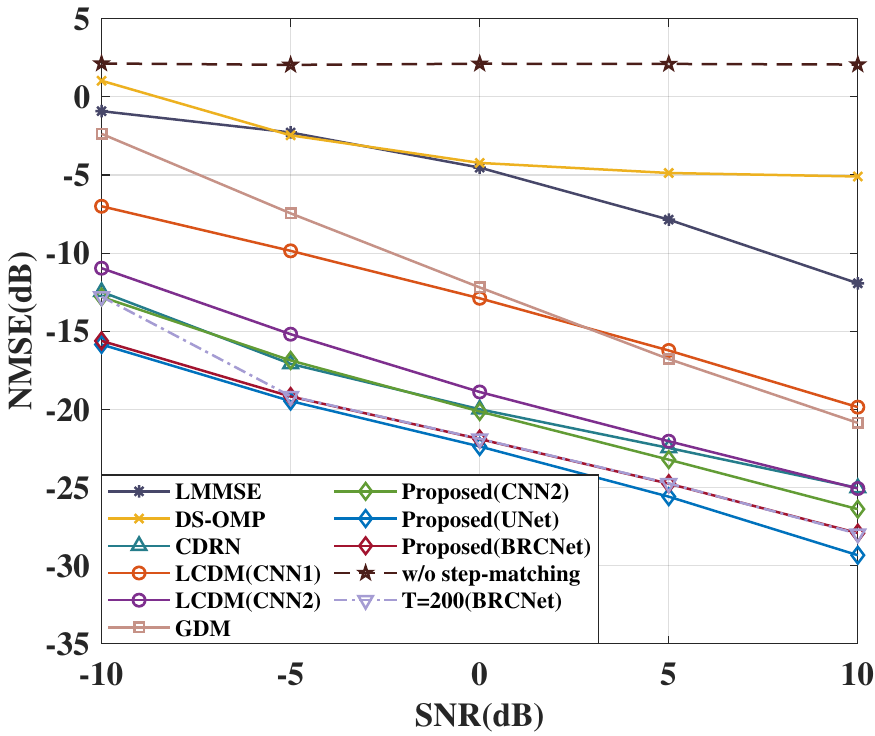}%
        \label{fig3:d}
    }
    \caption{NMSE performance versus SNR under different scenarios:(a) UMi, $f_c=28$GHz. (b) UMi, $f_c=73$GHz. (c) InH, $f_c=28$GHz. (d) InH, $f_c=73$GHz.}
    \label{fig3}
\end{figure*}

In simulations, the proposed method is compared with the following baselines: (i) the LMMSE estimator; (ii) DS-OMP~\cite{Wei2021}; (iii) CDRN~\cite{Liu2021}; (iv) LCDM (CNN1)~\cite{Fesl2024}, which is based on denoising diffusion probabilistic models (DDPMs) with a low-complexity CNN-based backbone; (v) LCDM (CNN2)~\cite{Fesl2024}, an enhanced version with an increased number of convolutional layers and feature channels to ensure a fair comparison; and (vi) GDM~\cite{Tong2024}, a method based on guided DMs for RIS channel estimation. For the proposed method, we implement three network architectures: CNN2, the U-Net architecture in~\cite{Wu2024}, and the proposed BRCNet.

Fig.~\ref{fig3} illustrates the NMSE performance under different SNR conditions. As an example, Fig.~\ref{fig3}\subref{fig3:d} (0 dB) is analyzed in detail. Unlike the direct-mapping approach of CDRN, the proposed method achieves superior performance through progressive denoising. Compared with LCDM (CNN1) and LCDM (CNN2), Proposed (BRCNet) achieves gains of 9.01 dB and 3.01 dB, respectively, demonstrating its superior effectiveness. By introducing the step-matching mechanism, our method better aligns with channel estimation and therefore outperforms GDM. Furthermore, the comparison between Proposed (CNN2) and Proposed (BRCNet) shows that the tailored BRCNet architecture is more effective than a simple CNN under the same parameter budget. Moreover, the proposed BRCNet achieves performance highly comparable to that of U-Net, with a performance gap of only 0.48 dB, while significantly reducing the number of parameters.

We further conduct an ablation study to evaluate the impact of different diffusion step numbers $T$ and the step-matching mechanism. By changing $T$ in the “Proposed (BRCNet)” from 1000 to 200, we obtain the “T=200 (BRCNet)” curve. The results show that $T=1000$ outperforms $T=200$ at low SNRs due to its finer noise-injection and denoising granularity, while the two settings yield similar performance at high SNRs, where only a few denoising steps are sufficient. We also remove the step-matching mechanism to obtain the “w/o step-matching” curve. Without step-matching, the sampling process starts from random Gaussian noise and does not utilize the received signal, resulting in SNR-insensitive performance. In contrast, with step-matching, sampling begins from the received signal, enabling effective denoising and yielding better performance that better aligns with the requirements of the channel estimation.

Table~\ref{Table1} compares the number of parameters and the number of multiply-accumulate operations (MACs) of different networks. Although CNN1 has fewer parameters, its limited capacity leads to inferior performance under complex channel conditions. CNN2 has a similar parameter count to that of BRCNet but still yields higher NMSE, highlighting the effectiveness of the proposed architecture. Notably, BRCNet achieves NMSE close to U-Net while using only 9.34\% of its parameters, offering an excellent trade-off among model complexity, computational overhead, and estimation accuracy.

\begin{table}[!t]
\caption{\textbf{ Parameters and Complexity Comparison}}
\label{Table1}
\centering
\begin{threeparttable} 
    \begin{tabular}{ccc}
    \toprule
    \textbf{Model}&\textbf{Parameters}&\textbf{MACs}\\
    \toprule  
    CNN1 & 55.03K & 57.32M \\
    CNN2 & 2.38M & 2.58G \\
    U-Net & 25.81M & 7.97G \\ 
    Proposed BRCNet & 2.41M & 2.18G \\ 
    \bottomrule
    \end{tabular}
\end{threeparttable}
\end{table}

\section{Conclusion}

In this letter, we propose a DMs-aided channel estimation method by formulating the estimation task as the reverse process of DMs, which adapts to varying noise levels. To reduce model complexity, we further design a parameter-efficient network, BRCNet, which achieves comparable performance with substantially fewer parameters. Simulation results demonstrate that the proposed method significantly outperforms baselines.

\bibliographystyle{IEEEtran_modified}
\bibliography{IEEEabrv,Reference}

\end{document}